\documentclass[a4paper]{elsart5p}
\usepackage[T1]{fontenc}
\usepackage[latin1]{inputenc}
\usepackage{array}
\usepackage{verbatim}
\usepackage{amsmath}
\usepackage{color}
\usepackage{graphicx}
\usepackage{amssymb}
\providecommand{\tabularnewline}{\\}

\allowdisplaybreaks

\usepackage{bm}

\usepackage{slashed}

\usepackage{url}

\usepackage{braket}

\newcommand{\ssst}[1]{\scriptscriptstyle{#1}}
\newcommand{\wt}[1]{\widetilde{#1}}

\newcommand{\mN}{m_{\ssst{N}}}

\newcommand{\vect}{\vec{\tau}}

\newcommand{\vecp}{\vec{\pi}}

\newcommand{\PV}{\slashed{P}}

\newcommand{\TV}{\slashed{T}}
\newcommand{\PVTV}{\slashed{P}\slashed{T}}

\usepackage[english]{babel}

\begin{document}
\begin{frontmatter}

\hfill \fbox{\parbox[t]{1.25in}{LA-UR-08-2293\\LLNL-JRNL-402812}}

\title{Nuclear Electric Dipole Moment of\ $^{3}\mathrm{He}$}
\author[lanl]{I. Stetcu}, 
\author[lanl,wisc,ornl]{C.-P. Liu}, 
\author[lanl]{J. L. Friar}, 
\author[lanl]{A. C. Hayes}, 
\author[llnl]{P. Navr\'atil}
\address[lanl]{Theoretical Division, Los Alamos National Laboratory, Los Alamos, NM 87545, U.S.A.} 
\address[wisc]{Department of Physics, University of Wisconsin-Madison, Madison, WI 53706, U.S.A.} 
\address[ornl]{Physics Division, Oak Ridge National Laboratory, Oak Ridge, TN 37831, U.S.A.} 
\address[llnl]{Lawrence Livermore National Laboratory, P.O. Box 808, L-414, CA 94551, U.S.A.}
\begin{abstract} 
A permanent electric dipole moment (EDM) of a physical system requires time-reversal ($T$) and parity ($P$) violation. Experimental programs are currently pushing the limits on EDMs in atoms, nuclei, and the neutron to regimes of fundamental theoretical interest. Here we calculate the magnitude of the $PT$-violating EDM of $^{3}$He and the expected sensitivity of such a measurement to the underlying $PT$-violating interactions. Assuming that the coupling constants are of comparable magnitude for $\pi$-, $\rho$-, and $\omega$-exchanges, we find that the pion-exchange contribution dominates. Our results suggest that a measurement of the $^{3}$He EDM is complementary to the planned neutron and deuteron experiments, and could provide a powerful constraint for the theoretical models of the pion-nucleon $PT$-violating interaction.  
\end{abstract}

\end{frontmatter}

\section{Introduction}

A permanent electric dipole moment (EDM) of a physical system would
indicate direct violation of time-reversal ($T$) and parity ($P$)
and thus $CP$ violation through the $CPT$ invariance. Presently
there are several experimental programs pushing the limits on EDMs
in atoms, nuclei, and the neutron to regimes of fundamental interest.
The Standard Model (SM) predicts values for the EDMs of these systems
that are too small to be detected in the foreseeable future, and hence
a measured nonzero EDM in any of these systems is an unambiguous signal
for a new source of $CP$ violation and for physics beyond the SM.
A new experimental scheme\ \cite{Khriplovich:1998zq,khrip2,Farley:2003wt,Semertzidis:2003iq,Orlov}
for measuring EDMs of nuclei (stripped of their atomic electrons)
in a magnetic storage ring suggests that the EDM of the deuteron could
be measured to an accuracy of better that 10$^{-27}$ $e$ cm\ \cite{Semertzidis:2003iq}.
Unlike searches for $CP$-violating moments of the nucleus through
measurements of atomic EDMs, a measurement for a stripped nucleus
would not suffer from a suppression of the signal through atomic Schiff
screening\ \cite{schiff}. For this reason, the latter could represent
about an order of magnitude better sensitivity to the underlying $CP$-violating
interaction than the present limit on the neutron EDM, $d_{n}$\ \cite{khrip2}.
Measurements using stripped nuclei in a magnetic storage ring are
best suited to nuclei with small magnetic anomaly, making $^{3}$He
an ideal candidate for a high precision measurement. Here we examine
the nuclear structure issues determining the EDM of $^{3}$He and
calculate the matrix elements of the relevant operators using the
no-core shell model\ \cite{Navratil:2000wwNavratil:2000gs} and
Podolsky's method for implementing second-order perturbation theory\ \cite{podolskyfriar}.
An approximate and incomplete calculation for the $^3$He dipole exists in
the literature \cite{Avishai:1986dw}, but here we present much more reliable calculations based 
on an exact solution of the three-body problem using several
potential models for the nucleon-nucleon (NN) interaction, complemented with three-body forces.

\section{Sources of Nuclear $P$,$T$ violation}

A nuclear EDM consists of contributions from the following sources:
(i) the intrinsic EDMs of the proton and neutron, $d_{p}$ and $d_{n}$;
(ii) the polarization effect caused by the $P$-,$T$-violating ($\PVTV$)
nuclear interaction, $H_{\PVTV}$; (iii) the two-body $\PVTV$ meson-exchange
charge operator appropriate for $H_{\PVTV}$.

The contribution due to nucleon EDMs, $D^{(1)}$, which is purely
one-body, can be easily evaluated by taking the matrix element \begin{align}
D^{(1)} & =\braket{\psi|\sum_{i=1}^{A}\,\frac{1}{2}\,\left[(d_{p}+d_{n})+(d_{p}-d_{n})\,\tau_{i}^{z}\right]\,\sigma_{i}^{z}|\psi}\,,\label{eq:1-body}\end{align}
 where $\ket{\psi}$ is the nuclear state that has the maximal magnetic
quantum number. In the particular case of interest in this paper,
$\ket{\psi}=\ket{0}$ is the ground state of $^{3}$He obtained by
the diagonalization of the $P$,$T$-conserving interaction.

In perturbation theory $H_{\PVTV}$ induces a parity admixture to
the nuclear state \begin{align}
\wt{\ket{0}} & =\sum_{n\neq0}\,\frac{1}{E_{0}-E_{n}}\,\ket{n}\braket{n|H_{\PVTV}|0}\,,\label{eq:mixture}\end{align}
 where $\ket{n}$ are eigenstates of energy $E_{n}$ and opposite
parity from $\ket{0}$, which are calculated with the $P$-,$T$-conserving
Hamiltonian. Hence, the polarization contribution $D^{(pol)}$ can
be simply calculated as \begin{align}
D^{(pol)} & =\bra{0}\,\hat{D}_{z}\wt{\ket{0}}+\mbox{c.c.}\,,\label{eq:pol}\end{align}
 where \begin{align}
\hat{D}_{z} & =\frac{e}{2}\,\sum_{i=1}^{A}\,(1+\tau_{i}^{z})\, z_{i}\end{align}
 is the usual dipole operator projected in the $z$-direction.

The contribution due to exchange charge, $D^{(ex)}$, is typically
at the order of $(v/c)^{2}$, and explicitly evaluated to be just
a few percent of the polarization contribution for the deuteron case\ \cite{Liu:2004tq};
we therefore ignore it and approximate the full two-body contribution,
$D^{(2)}$, solely by the polarization term \begin{align}
D^{(2)} & =D^{(pol)}+D^{(ex)}\cong D^{(pol)}\,.\label{eq:2-body}\end{align}

Our calculation of the EDM of $^{3}$He therefore requires knowledge
of both the individual EDMs of the nucleons and the $\PVTV$ nuclear
force. These very different quantities can only be related if some
understanding exists of both the origin of the symmetry violation
and its expression in strong-interaction observables. Constructing
an effective field theory (EFT) that incorporates the symmetry violation,
as well as the dynamics underlying the usual strong-interaction physics
in nucleons and nuclei, provides a suitable framework. Chiral Perturbation
Theory ($\chi$PT) supplemented with a knowledge of the symmetry violation
would be the appropriate EFT. To date only a single such calculation
exists \cite{Hockings:2005cn}, and it was applied to the one-nucleon
sector, although further effort is underway \cite{-PVTV-EFT,Bira:PC}.
The symmetry violation in that calculation was taken from the QCD
$\bar{\theta}$ term, which leads to an isoscalar $\PVTV$ pion-nucleon
interaction in leading order, unlike the most general case that includes
an isovector and an isotensor term, as well\ \cite{Barton:1969gi}.
The non-analytic  parts of the pion-loop diagrams\ \cite{Liu:2004tq,Hockings:2005cn,Crewther:1979pi,He:1989xj} that generate nucleon EDMs then provide an appropriate  estimate of the EDMs of individual nucleons. These contributions are expected to dominate in the chiral limit\cite{Crewther:1979pi}.

In the absence of a $\chi$PT calculation of $H_{\PVTV}$ we revert
to a conventional formulation in terms of a one-meson-exchange model.
Including $\pi$-, $\rho$-, and $\omega$-meson exchanges,\ %
\footnote{Other mechanisms not included here are $\eta$-exchange and two-$\pi$-exchange,
etc.%
} the interaction is given by (see Refs.\ \cite{Liu:2004tq,Haxton:1983dq,Herczeg:1987gp,Gudkov:1993yc,Towner:1994qe}):
\begin{align}
 & H_{\PVTV}(\bm r) =\frac{1}{2\,\mN}\,\bigg\{\bm\sigma_{-}\cdot\bm\nabla\left(-\bar{G}_{\omega}^{0}\, y_{\omega}(r)\right)\nonumber \\
 & +\bm\tau_{1}\cdot\bm\tau_{2}\,\bm\sigma_{-}\cdot\bm\nabla\left(\bar{G}_{\pi}^{0}\, y_{\pi}(r)-\bar{G}_{\rho}^{0}\, y_{\rho}(r)\right)\nonumber \\
 & +\frac{\tau_{+}^{z}}{2}\,\bm\sigma_{-}\cdot\bm\nabla\left(\bar{G}_{\pi}^{1}\, y_{\pi}(r)-\bar{G}_{\rho}^{1}\, y_{\rho}(r)-\bar{G}_{\omega}^{1}\, y_{\omega}(r)\right)\nonumber \\
 & +\frac{\tau_{-}^{z}}{2}\,\bm\sigma_{+}\cdot\bm\nabla\left(\bar{G}_{\pi}^{1}\, y_{\pi}(r)+\bar{G}_{\rho}^{1}\, y_{\rho}(r)-\bar{G}_{\omega}^{1}\, y_{\omega}(r)\right)\nonumber \\
 & +(3\,\tau_{1}^{z}\,\tau_{2}^{z}-\bm\tau_{1}\cdot\bm\tau_{2})\,\bm\sigma_{-}\cdot\bm\nabla\left(\bar{G}_{\pi}^{2}\, y_{\pi}(r)-\bar{G}_{\rho}^{2}\, y_{\rho}(r)\right)\bigg\}\,,\label{eq:HPVTV-ex}\end{align}
where $\mN$ is the nucleon mass, $\bar{G}_{x}^{T}$ is defined as the product of a $\PVTV$
x-meson--nucleon coupling $\bar{g}_{x}^{T}$ (with $T$ referring
to the isospin) and its associated strong one, $g_{x\ssst{NN}}$
(e.g., $\bar{G}_{\pi}^{0}=g_{\pi\ssst{NN}}\,\bar{g}_{\pi}^{0}$,where the interaction Lagrangian corresponding to these coupling constants is ${\cal{L}} = \bar{N} [i  g_{\pi\ssst{NN}} \gamma_5+  \bar{g}_{\pi}^{0}] \vect \cdot \vecp N$), 
$y_{x}(r)=e^{-m_{x}\, r}/(4\,\pi\, r)$ is the Yukawa
function with a range determined by the mass of the exchanged $x$-meson,
$\vec r=\vec r_1 -\vec r_2$, $\vec \sigma_\pm=\vec \sigma_1\pm \vec \sigma_2$, and similarly for $\vec \tau_\pm$. Unless the symmetry associated with the specific way that $P,T$ violation is generated suppresses some of the couplings, one expects (by naturalness) that these $\PVTV$ meson--nucleon couplings are of similar magnitude, and this is roughly confirmed by a QCD sum
rule calculation\ \cite{Pospelov:2001ys}.\ %
\footnote{In Ref.\ \cite{Pospelov:2001ys}, $\bar{g}_{\rho,\omega}$ are defined differently; for conversion, see Ref.\ \cite{Liu:2004tq}%
} We note, however,  that in the (purely isoscalar) $\bar{\theta}$-term model 
of Ref.\ \cite{Hockings:2005cn} the coupling constants $\bar{G}_{\pi}^{1}$
and $\bar{G}_{\pi}^{2}$ vanish, and the coupling constants for the
short-range operators are very small compared to the pion one\ \cite{Bira:PC}.

Because $H_{\PVTV}$ violates parity and $^{3}$He is (largely) an
$S$-wave nucleus, the matrix elements that define the EDM (see below)
mostly involve $S$- to $P$-wave transitions. This has the combined effect
of suppressing the short-range contributions and enhancing the long-range
(i.e., pion) contribution, irrespective of the detailed nature of
the force. Combined with the consideration that the short-range parameters
($\bar{G}_{\rho,\omega}$) are not much larger than the pion ones,
one can roughly expect the dominance of pion exchange.

\section{$^{3}\mathrm{He}$ in the \textit{ab initio} No-Core Shell Model}

We solve the three-body problem in an \textit{ab initio} no-core shell
model (NCSM) framework \cite{Navratil:2000wwNavratil:2000gs}. The
ground-state wave function is obtained by a direct diagonalization
of an effective Hamiltonian in a truncated harmonic oscillator (HO)
basis constructed in relative coordinates, as described in Ref.\ \cite{Navratil:1999pw}.
High-precision NN interactions, such as the local Argonne
$v_{18}$\ \cite{Wiringa:1995Pieper:2001AR} and the non-local charge-dependent
(CD) Bonn potential\ \cite{Machleidt:1995km} interactions, are used
to derive an effective interaction in each model space via a unitary
transformation\ \cite{Okubo:1954DaProvidencia:1964Suzuki:1980}
in a two-body cluster approximation. The Coulomb interaction between
protons is also taken into account.

In addition to the phenomenological NN interaction models cited above,
we consider two- and three-body interactions derived from EFT. In
a recent work\ \cite{Navratil:2007} using the 
NCSM, the presently available NN potential at N$^{3}$LO\ \cite{N3LO}
and the three-nucleon (NNN) interaction at N$^{2}$LO\ \cite{vanKolck:1994,Epelbaum:2002}
have been applied to the calculation of various properties of $s$-
and $p$-shell nuclei. In that study a preferred choice of the two
NNN low-energy constants, $c_{D}$ and $c_{E}$, was found (and the
fundamental importance of the chiral EFT NNN interaction was demonstrated)
by reproducing the structure of mid-$p$-shell nuclei. (Note that
these interactions are fitted only for a momentum cutoff of 500 MeV,
and therefore we are not able at this time to demonstrate a running
of the observables with the cutoff.) This Hamiltonian was then used
to calculate microscopically the photo-absorption cross section of
$^{4}$He\ \cite{Quaglioni:2007eg}, while the full technical details
on the local chiral EFT NNN interaction that was used were given in
Ref.\ \cite{Navratil_FBS}. We use an identical Hamiltonian in the
present work, and we compare its predictions against the phenomenological
potentials.

In the NCSM the basis states are constructed using HO wave functions.
Hence, all the calculations involve two parameters: the HO frequency
$\Omega$ and $N_{max}$, the number of oscillator quanta included
in the calculation. At large enough $N_{max}$, the results become
independent of the frequency, although the rate of convergence depends
on $\Omega$. Thus, for short-range operators, one can expect a faster
convergence for larger values of $\Omega$, as the characteristic
length of the HO is $b=1/\sqrt{\mN\,\Omega}$. The convergence also
depends upon the $P$-,$T$-conserving interaction, $H_{0}$, used
to solve the three-body problem. Thus the results obtained with Argonne
$v_{18}$ show the slowest convergence, because the NN interaction
has a more strongly repulsive core than the interactions obtained
from EFT, which have faster convergence rates.

The nucleonic contribution $D^{(1)}$ in Eq.\ (\ref{eq:1-body})
involves only $H_{0}$, and is easily calculated once the three-body
problem is solved. We therefore concentrate on the part involving
$H_{\PVTV}$.

Equation\ (\ref{eq:mixture}) suggests that in order to calculate
the dipole moment one needs to obtain high-accuracy excited states
of $^{3}$He in the continuum, an extremely difficult task in a NCSM
framework, where the basis states are constructed using bound-state
wave functions. The most straightforward technique for evaluating
Eq.\ (\ref{eq:mixture}) is to use Podolsky's method\ \cite{podolskyfriar},
in which $\wt{\ket{0}}$ is obtained as the solution of the Schr\"odinger
equation with an inhomogeneous term 
\begin{equation}
(E_{0}-H_{0})\,\wt{\ket{0}}=H_{\PVTV}\,\ket{0}\,.
\label{nonhomeq}
\end{equation}
 The exceptionally nice feature of this method is that continuum states
do not have to be \textit{explicitly} calculated (they are, of course,
implicitly included). In this sense the technique is a relatively
simple extension of bound-state methods, which have been well studied
and are robust. Moreover, in this approach the convergence of the
EDM reduces to a large degree to the issue of the convergence of the
ground state.

We express the solution of Eq.\ (\ref{nonhomeq}) as a superposition
of a handful of vectors generated using the Lanczos algorithm\ \cite{LanczAlg1950LanczAlg}.
Indeed, one can show that if we start with the inhomogeneous part
of Eq.\ (\ref{nonhomeq}) as the starting Lanczos vector $|v_{1}\rangle=H_{\PV\TV}\,|0\rangle$,
the solution becomes\ \cite{LanczGreen}

\begin{equation}
\wt{\ket{0}}\approx\sum_{k}\, g_{k}(E_{0})\,|v_{k}\rangle\,,\end{equation}
 where the summation over the index $k$ runs over a finite and usually
small number of iterations. The coefficients $g_{k}(E)$ are easily
obtained using finite continued fractions\ \cite{LanczCoef}.

We alter this approach in practice for efficiency reasons. Because
Eq.\ (\ref{eq:pol}) is symmetrical in $\hat{D}_{z}$ and $H_{\PV\TV}$,
we are free to choose $|v_{1}\rangle=\hat{D}_{z}\,|0\rangle$ as the
starting vector. This allows us to isolate the two isospin contributions
for $H_{\PV\TV}$ in each run. Once we compute a second vector, $|v\rangle=H_{\PV\TV}^{\dagger}\,|0\rangle$,
the polarization contribution to the EDM is finally evaluated as \begin{equation}
D^{(pol)}=2\,\sum_{k}\, g_{k}(E_{0})\,\langle v|v_{k}\rangle\,.\label{dipoleLanczos}\end{equation}
 (We have verified that the altered approach gives the same results
as the original one.) As a particular test case we have considered
the electric polarizability

\begin{equation}
\alpha_{E}=\frac{1}{2\pi^{2}}\,\int\, d\omega\,\frac{\sigma(\omega)}{\omega^{2}}=2\,\alpha\,\sum_{n}\,\frac{\langle0|\hat{D}_{z}|n\rangle\langle n|\hat{D}_{z}|0\rangle}{E_{n}-E_{0}}\end{equation}
 (where $\alpha$ is the fine structure constant), which reduces Eq.\ (\ref{dipoleLanczos})
to $\alpha_{E}=-2\,\alpha\, g_{1}(E_{0})\,\langle v_{1}|v_{1}\rangle$.
We estimate that the electric polarizability of the $^{3}$He nucleus
is 0.183\ fm$^{3}$ for the Argonne $v_{18}$ potential, compared
with 0.159\ fm$^{3}$ reported in Ref.\ \cite{He3alphaE} for the
same interaction. The $15\%$ discrepancy is most likely the result
of a difference in the theoretical approaches, as the result reported
in Ref.\ \cite{He3alphaE} involves a matching of the ground-state
energy to experiment (i.e., 7.72 MeV), although the calculation gives
6.88\ MeV binding\ \cite{gsHe3energyAV18} in the absence of three-body
forces (our converged binding energy for Argonne $v_{18}$ is 6.92\ MeV).
Since the electric polarizability scales roughly with the inverse
of the square of the binding energy, the discrepancy between the two
results is reasonable. Moreover, we have made the additional check
of the Levinger-Bethe sum rule\ \cite{BetheSR}, which in the case
of tritium relates the total dipole strength to the charge radius,
and we found it to be satisfied in all model spaces to a precision
better than $10^{-5}$. Finally, the $^{3}$He polarizability calculated
using the two- and three-body chiral interactions is 0.148\ fm$^{3}$,
compared with 0.145\ fm$^{3}$ with Argonne $v_{18}$ and Urbana
IX\ \cite{He3alphaE} two- and three-body forces. In both cases excellent
agreement with the experimental binding energy is achieved.

In a consistent approach the same transformation used to obtain the
effective interaction should be used to construct the effective operators
in truncated spaces. While this has been done in the past for general
one- and two-body operators\ \cite{Stetcu:2004wh},
such an approach is very cumbersome for the present investigation
because both the dipole transition operator and $H_{\PVTV}$ change
the parity of the states. We have therefore chosen not to renormalize
the operators involved, except for the $P$-,$T$-conserving Hamiltonian.
This problem is largely overcome by the fact that long-range operators
(like the dipole) have been found to be insensitive to the renormalization
in the two-body cluster approximation\ \cite{Stetcu:2004wh},
which is the level of truncation for the effective interaction. We
also point out that since $H_{\PVTV}$ has short range, one can
expect that the renormalization of this operator would improve the
convergence pattern, especially for small HO frequencies. As with all
operators, the effect of the renormalization decreases as the size
of the model space increases, so that in large model spaces (like
the ones in the present calculation) this effect can be safely neglected
and good convergence of observables is achieved.

\section{Results and Discussions}

We start the discussion of our results with the one-body contribution
to the EDM of $^{3}$He. In Table\ \ref{table:oneb}, we summarize
the isoscalar and isovector contribution to $D^{(1)}$, which are
decomposed into contributions proportional to their respective coupling
constants ($d_{p}+d_{n}$ for isoscalar, and $d_{p}-d_{n}$ for isovector).
All interactions give similar results, with only the Argonne $v_{18}$
result deviating more significantly from the others, albeit by less
than 6\%. We note that the coefficients in the upper and lower rows
in Table\ \ref{table:oneb} would be either 1/2 or -1/2 if the nuclear
forces between each pair of nucleons were taken to be equal (viz.,
the $SU(4)$ limit, which implies that the neutron carries all of
the nuclear spin).

\begin{table}[h]

\caption{The nucleonic contribution (in $e$ fm) to the $^{3}$He EDM for different
potential models. We decompose our results into contributions proportional
to the nucleon isoscalar ($d_{p}+d_{n}$) and isovector ($d_{p}-d_{n}$)
EDMs.}

\ \label{table:oneb}

\centering

\begin{tabular}{c|r|r|rr}
\hline 
 & CD Bonn  & $v_{18}$  & \multicolumn{2}{c}{EFT}\tabularnewline
 &  &  & NN  & NN+NNN \tabularnewline
\hline 
$d_{p}+d_{n}$  & $0.430$  & $0.415$  & $0.437$  & $0.433$ \tabularnewline
$d_{p}-d_{n}$  & $-0.467$  & $-0.462$  & $-0.468$  & $-0.468$ \tabularnewline
\hline
\end{tabular}
\end{table}

\begin{figure}[h]
\centering\includegraphics[clip,scale=0.87]{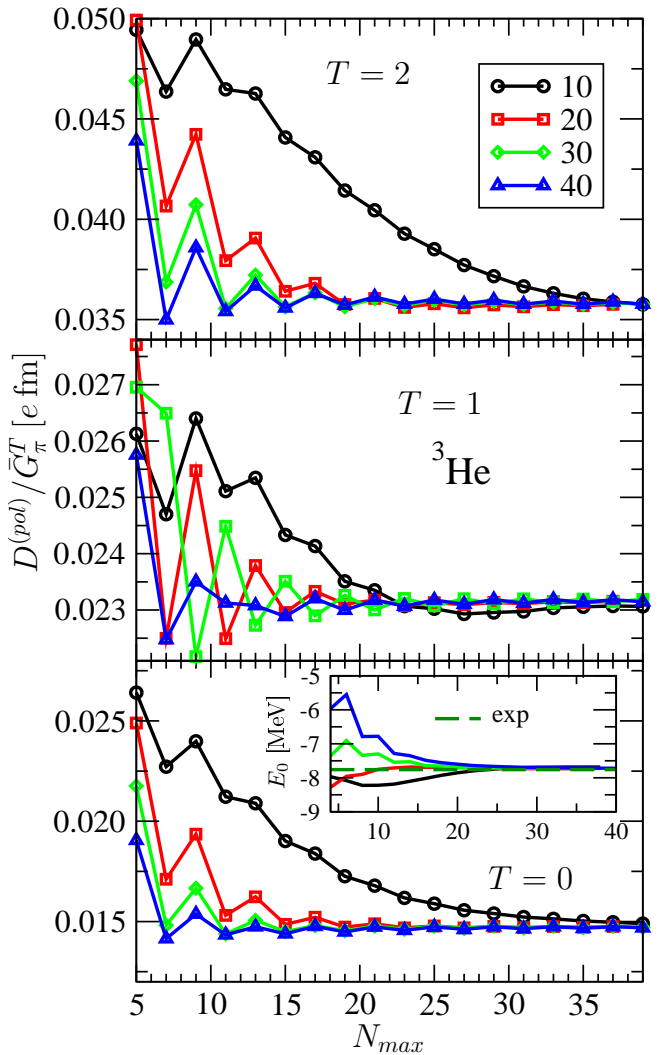}

\caption{Isoscalar, isovector, and isotensor pion-exchange contributions to
the $H_{\PVTV}$-induced EDM in $^{3}$He. We show four different
frequencies in each case: $\Omega=10$\ MeV (circles), $\Omega=20$\ MeV
(squares), $\Omega=30$\ MeV (diamonds), and $\Omega=40$\ MeV (triangles).
In the insert, we present the convergence of the ground-state energy,
which in the limit of large $N_{max}$ approaches the experimental
value (dashed line). Both NN and NNN EFT interactions have been used
for diagonalization.}

\label{fig:Dpol} 
\end{figure}

In Fig.\ \ref{fig:Dpol} we present for four HO frequencies the running
with $N_{max}$ of the EDM induced by the pion-exchange part of $H_{\PVTV}$.
Two- and three-body EFT interactions have been used for this calculation,
in order to obtain an accurate description of the ground-state of
the three-body system. For the nuclear EDM we mix two types of operators:
$H_{\PV\TV}$, which is short range, and $\hat{D}_{z}$, which is
long range. The convergence pattern is therefore not as straightforward
as presented earlier in the discussion of convergence properties of general
operators. The short-range
part dominates the convergence pattern, and we thus observe faster
convergence for larger frequencies (smaller HO parameter length).
This behavior is opposite to the convergence in the case of the electric
polarizability, where we found faster convergence for smaller frequencies
as expected for a long range operator.
Nevertheless, just as in Fig.\ \ref{fig:Dpol}, the results become
independent of the frequency at large $N_{max}$. Note in the insert
the convergence behavior of the ground-state energy, which converges
to the experimental value already at $N_{max}\approx22$ for most
frequencies presented in the figure.

Similar convergence patterns can be observed for the other meson exchanges
as well as other potential models. In Table\ \ref{table:HPTedm}
we summarize these results.

\begin{table*}[t]

\caption{The polarization contribution to $^{3}$He EDM (in units of $e$ fm),
decomposed as coefficients of $\bar{G}_{x}^{T}$, where $x$ stands
for $\pi$, $\rho$, or $\omega$ meson exchanges.}

\ \label{table:HPTedm}

\centering
\begin{tabular*}{\textwidth}{@{\extracolsep{\fill}} l*{4}{r}|*{4}{r}|*{4}{r}}
\hline
\hline
& \multicolumn{4}{c|}{$\pi$} & \multicolumn{4}{c|} {$\rho$}  & \multicolumn{4}{c} {$\omega$}  \\
\cline{2-13}  & CD Bonn & $v_{18}$  & \multicolumn{2}{c|}{EFT}  & CD Bonn &$v_{18}$ & \multicolumn{2}{c|}{EFT} & CD Bonn & $v_{18}$  & \multicolumn{2}{c}{EFT}\\
&  &  &  NN & NN+NNN &  &  & NN & NN+NNN & & & NN &NN+NNN\\
\hline
$\bar G_{x}^0$& $0.013$ & $0.012$ & $0.015$ & $0.015$ & $-0.0012$ & $-0.0006$ & $-0.0012$ & $-0.0013$& $0.0008$ & $0.0005$  & $0.0009$ & $0.0007$\\
$\bar G_{x}^1$ & $0.022$ & $0.022$ & $0.023$ & $0.023$ & $0.0011$ & $0.0009$ & $0.0013$ & $0.0012$ & $-0.0011$ & $-0.0011$ & $-0.0017$ & $-0.0018$\\
$\bar G_{x}^2$ & $0.035$ & $0.034$ & $0.037$ & $0.036$ & $-0.0019$ & $-0.0015$ & $-0.0028$ & $-0.0027$ &- &- & - & -\\
\hline
\hline
\end{tabular*}

\end{table*}

For pion exchange all potential models give basically the same result,
as the long-range part ($r\gtrsim1/m_{\pi}$) of the $^{3}\mbox{He}$
wave function shows negligible model dependence. It is interesting
to note the effect of the three-body force by comparing the results
with and without NNN interactions. When only the NN EFT interaction
is used, the binding energy is underestimated by about 500\ keV.
Therefore, since the ground-state energy is in the denominator of
Eq.\ (\ref{eq:mixture}), one could naively expect that introducing
the three-body forces (which increase the binding) decreases $D^{(pol)}$.
Instead we obtain nearly the same results for both isoscalar and isovector
contributions. This implies that the NNN interaction reshuffles the
strength to compensate for the change in binding energy, most likely
at low energies. This is not a surprise, because it was already found
previously that the main effect of the three-body forces for the dipole
response is an attenuation of the peak region at low energies both
in the three-\ \cite{He3alphaE} and four-body\ \cite{Quaglioni:2007eg}
systems.

In the isoscalar pion-exchange channel our result ($0.015$) is about
$50\%$ larger than an existing work\ \cite{Avishai:1986dw}, which
yielded $0.010$. Since the Reid soft core NN potential used in Ref.\ \cite{Avishai:1986dw}
has a much more repulsive core than
even Argonne $v_{18}$, that isoscalar contribution to the EDM is in line with our findings. 
Moreover, the approximate solution of the three-body
problem (as opposed to the current work, in which the calculation is exact) can
induce uncontrolled uncertainties. Finally, their calculation of the isovector term
did not exhaust all the possible spin-isospin combinations, while the isotensor
contribution was not computed. 

For $\rho$- and $\omega$-exchanges one immediately sees that their
corresponding coefficients are at most $10\%$ of the pion-exchange
ones, because only the short-range wave function ($r\lesssim1/m_{\rho,\omega}$)
contributes substantially. Sensitivity of the short-range $\PVTV$ potentials
to the short-range model dependence of the wave functions produces
matrix-element variations as large
as $50\%$ for some channels. While a detailed explanation for the
model dependence is too intricate to be disentangled, one can roughly
see the general trend that the calculation using Argonne $v_{18}$
gives consistently smaller results than ones using CD Bonn and chiral
EFT, as Argonne $v_{18}$ has a harder core than the other two. The
behavior of short-range $\PVTV$ nuclear potentials in $\chi$PT strongly suggests
that the short-range coupling constants are no larger than the pion
ones, and may be significantly smaller. If this holds, pion-exchange
will produce the dominant contribution to nuclear EDMs, with the heavy-meson
exchanges (the short-range interaction) giving roughly a $10\%$ correction
(or less) to the pion-exchange contribution. This suppression due
to $P$-wave intermediate nuclear states was discussed earlier and is in accord
with calculations in heavier systems \cite{Towner:1994qe}.

Assuming the dominance of pion exchange, $D^{(2)}$ has an almost
model-independent expression \begin{align}
D^{(2)} & \approx(0.015\,\bar{G}_{\pi}^{0}+0.023\,\bar{G}_{\pi}^{1}+0.036\,\bar{G}_{\pi}^{2})\, e\,\mbox{fm}.\end{align}
 The single-nucleon EDMs can be estimated using the non-analytic term
that results from calculating the one-pion loop diagram, which dominates
in the chiral limit (see, for example, Refs.\ \cite{Liu:2004tq,Hockings:2005cn,Barton:1969gi,Crewther:1979pi,He:1989xj})

\begin{align}
d_{\stackrel{{\scriptstyle p}}{n}} & \approx\mp\frac{e}{4\,\pi^{2}\,\mN}\,(\bar{G}_{\pi}^{0}-\bar{G}_{\pi}^{2})\,\ln\left(\frac{\mN}{m_{\pi}}\right)\,,\label{eq:nucleon-EDM}\end{align}
where $e$ is the proton charge. Folding this result into $D^{(1)}$ and using the physical nucleon
and pion masses ($\ln(\mN/m_{\pi})\approx1.90$) we get \begin{align}
D^{(1)} & \approx0.009\,(\bar{G}_{\pi}^{0}-\bar{G}_{\pi}^{2})\, e\,\mbox{fm}\,.\end{align}
 The total EDM of $^{3}\mbox{He}$ is therefore estimated to be \begin{align}
D & =D^{(1)}+D^{(2)}\nonumber \\
 & =(0.024\,\bar{G}_{\pi}^{0}+0.023\,\bar{G}_{\pi}^{1}+0.027\bar{G}_{\pi}^{2})\, e\,\mbox{fm}\,.\end{align}

Calculating the EDMs of the neutron and deuteron\ \cite{Liu:2004tq}
using Eq.\ (\ref{eq:nucleon-EDM}) and assuming pion-exchange dominance,
one can see from Table\ \ref{tab:comparison} that an EDM measurement
in $^{3}\mbox{He}$ is complementary to the former two.  Assuming that similar sensitivities
can be reached in these three measurements, the $\PVTV$ pion-nucleon
coupling constants could be well-constrained if the assumption of
pion-exchange dominance holds.

\begin{table}[h]

\caption{Complementarity of the $^3$He EDM to the neutron and deuteron EDMs. 
We present the theoretical estimations (in units of $e$ fm) of neutron, deuteron, and $^{3}\mbox{He}$
decomposed into contributions proportional to $\bar{G}_{\pi}^{0,1,2}$,
while assuming the dominance of pion-exchange forces in $H_{\PV\TV}$
and estimating nucleon EDMs via pion loops.\ \label{tab:comparison}}

\centering\begin{tabular}{>{\centering}p{0.2\columnwidth}>{\raggedleft}p{0.2\columnwidth}>{\raggedleft}p{0.2\columnwidth}>{\raggedleft}p{0.2\columnwidth}}
 & $\bar{G}_{\pi}^{0}$  & $\bar{G}_{\pi}^{1}$  & $\bar{G}_{\pi}^{2}$\tabularnewline
\hline
\hline 
neutron  & $0.010$  & $0.000$  & $-0.010$\tabularnewline
deuteron  & $0.000$  & $0.015$  & $0.000$\tabularnewline
$^{3}\mbox{He}$  & $0.024$  & $0.023$  & $0.027$\tabularnewline
\hline
\end{tabular}
\end{table}

\section{Summary}

In this paper, we have calculated the nuclear EDM of $^{3}$He, which
arises from the intrinsic EDMs of nucleons and the $P$-,$T$-violating
nucleon-nucleon interaction. Several potential models for the $P$-,$T$-conserving
nuclear interaction (including the latest-generation NN and NNN chiral
EFT forces) have been used in order to obtain the solution to the
nuclear three-body problem. The results obtained with these potential
models agree within 25\% in the $\PVTV$ pion-exchange sector. Though
larger spreads in $\PVTV$ $\rho$- and $\omega$-exchanges are found
(as the results sensitively depend on the wave functions at short
range), we expect them to be non-essential as pion-exchange completely
dominates the observable (unless the $\PVTV$ parameters associated
with heavy-meson exchanges are significantly larger than the ones
for pion exchange, which is not expected). We further demonstrate
that a measurement of the $^{3}\mbox{He}$ EDM would be complementary
to those of the neutron and deuteron, and in combination they can
be used to put stringent constraints on the three $P$-,$T$-violating
pion--nucleon coupling constants. We therefore strongly encourage
experimentalists to consider such a $^{3}$He measurement in a storage
ring, in addition to the existing deuteron proposal\ \cite{Semertzidis:2003iq}.

\begin{ack}

I.S. thanks W. Leidemann, S. Quaglioni and S. Bacca for useful discussions.
J.L.F. greatly appreciates insights provided by P. Herczeg and U.
van Kolck. The work of I.S., C.P.L., J.L.F. and A.C.H. was performed under the auspices of the U. S.  
DOE. C.P.L. acknowledges partial support from the Wisconsin Alumni Research Foundation  and
U.S. DOE under contract Nos. DE-FG02-08ER41531 (UWisc) and
DE-AC05-00OR22725 (ORNL). Prepared by LLNL under contract No. DE-AC52-07NA27344. Support from the 
U.S. DOE/SC/NP (Work Proposal No. SCW0498), and from the U. S. Department of Energy Grant 
DE-FG02-87ER40371 is acknowledged. 
\end{ack}


\end{document}